\documentstyle[11pt,multicol,emulateapj,psfig]{article}

\begin{document}


\title{A Further Study of Relative longitude shift of Pulsar Beams}

\author{R.X. Xu, J.W. Xu, G. J. Qiao\\
        CAS-PKU joint Beijing Astrophysical Center
        and Department of Astronomy, \\
        Peking University, Beijing 100871, China}

\altaffiltext{1}{email: rxxu@bac.pku.edu.cn, xjw@bac.pku.edu.cn,
gjn@pku.edu.cn}

\altaffiltext{2}{This work is supported by NSFC (No. 19803001),
by the Special Funds for Major State Basic Research Project, and
by Doctoral Program Foundation of Institution of Higher Education
in China.}

\begin{abstract}

It is of great important to study pulsar beam shapes if we are
concerned with emission theories and pulsar birth rate. Both
observations$^{[\cite{r83,r93}]}$ and/or the ICS
model$^{[\cite{ql98}]}$ show that different emission components
are emitted from different heights. The relative longitude phase
shifts due to different heights of emission components and to the
toroidal velocity of electron are considered in this paper.
Several possible observational features arising from the phase
shift effects are presented. The emission beams may not have
circular cross sections although the emission regions may be
symmetric with respect to the magnetic axes.

\end{abstract}


\section{Introduction}

Soon after the discovery of Pulsars, the rotating vector model
(RVM) was proposed to well explain the linear polarization
characteristics of pulsar radio emission, in which Radhakrishnan
and Cooke$^{[\cite{rc69}]}$ assumed that relativistic particles
stream out along dipolar magnetic fields of pulsars and the
polarization position angle of emitted radiation from these
particles is correlated with the magnetic curvature. Theoretical
extensions of the RVM have appeared in literatures (e.g.,
Ferguson$^{[\cite{ferg76}]}$, Blaskiewicz et.
al.$^{[\cite{mcw91}]}$, Hibschman \& Arons$^{[\cite{ha00}]}$),
where some detailed physical factors, such as the special
relativistic effects, the polar-cap current flow, have been
included.

However the RVM fails in attempting to explain the position angle
jumps observed in mean pulses as well as in individual
pulses$^{[\cite{sti84}]}$. This leads to the assumption that the
radio beams of pulsars contain two nearly orthogonal polarization
modes, the superposition of which results in the non-S-shaped
position angle variation in mean pulses and two $\sim 90^{\rm o}$
separated distributions of position angles in individual pulses
(see the paper by Mckinnon \& Stinebring$^{[\cite{ms98}]}$ and
references therein). Nevertheless, there is still a possibility
that modified RVM, {\em without}
inclusion of the orthogonal modes, can reproduce the observed
position angle ``jumps'' if we consider the linear depolarization
of pulsar beams and the observational uncertainties originated
from observation polarimeters$^{[\cite{xqh97,xq00}]}$.
In view of the observational fact that radio emission of pulsars
may have three components (core and two cones) which are emitted
from different heights above pulsar surface$^{[\cite{r83}]}$, Xu
et al.$^{[\cite{xqh97}]}$ found that the position angles of mean
pulses jump at certain longitudes where the linear polarization
is approximately zero if the retardation effect, which causes the
beam centers to be shifted between each other, can significantly
conduce toward the linear depolarization.
In this paper, we further study this issue arising from different
emission heights which are calculated according to the inverse
Compton scattering model$^{[\cite{ql98}]}$ in which core and
conal components can be naturally reproduced.

Besides the polarization effect discussed above, different
emission heights may also have notable consequence to beam shapes.
Many authors investigated observationally the pulsar beam shapes,
the previous results of which are summarized below.
Jones$^{[\cite{j80}]}$ find that the latitudinal radius of beam
cross section could be 2.5 times the longitudinal one. Narayan \&
Vivekanand$^{[\cite{nv83}]}$ modify this ratio to be 3.0. Lyne \&
Manchester's$^{[\cite{lm88}]}$ conclusion is that the shape of
the emission beam is approximately circular.
Biggs$^{[\cite{b90}]}$ proposes that there may be some meridional
compression. Wu \& Shen$^{[\cite{ws88}]}$ find that, for pulsars
with short period, the radius of the emission beam at latitudinal
direction is larger. However, the latitudinal radius is smaller
for pulsars with longer period. Rankin, therefore, change her
morphological beam shape from an elliptical cross
section$^{[\cite{r83}]}$ to a circular one$^{[\cite{r93}]}$. Our
numerical results in section 3 may be checked by such
observational study in the future.
We find that the emission beams of pulsars with small rotation
periods do not have circular cross sections even if the emission
regions are symmetric with respect to the magnetic axes if we
take into account the toroidal velocity due to rotation.

This paper is organized as below. After inspecting various
aspects of the retardation effect caused by different emission
heights and their possible observational consequences in section
2, we calculate the emission beams from relativistic particles
with poloidal (along magnetic field lines) and toroidal
(perpendicular to field lines ) velocities in section 3. A
conclusion, based on which we can know the phase shift between
two emission units at arbitrary positions, is proved
geometrically in Appendix. The paper is summarized in section 4.

\section{Relative longitude shift due to different heights}

It is generally believed that pulsars' radio emission is
originated from the region near the last open field lines. We
assume in this section that the particles in the
last-open-field-line region, which are responsible to observed
radiation, have only poloidal velocities along the filed lines
(i.e., the toroidal velocity is neglected), and are accelerated
symmetrically respect to the magnetic axis. The emission region
of each beam component is thus also symmetric respect to the
axis. According to Appendix, the phase shifts of all emission
elements of a certain beam component is the same since the values
${\bf r} \cdot {\bf n}_0$ of those emission elements are the
same, where ${\bf r}$ is the vector position of an element, ${\bf
n}_0$ the emission direction. Therefore the emission beams should
have circular cross sections in this case.

\subsection{Emission heights and beam shift}

We know from Appendix that an emission element at vector position
${\bf r}$ is equivalent to a virtual one whose longitude phase
shifts to an earlier value $\delta\phi (r) = \Omega {\bf r} \cdot
{\bf n}_0/c$. We now calculate $\delta\phi (r)$ as a function of
$r$, based on the dipole geometry of magnetic field configuration.

For a dipole field line, we have
\begin{equation}
r=\lambda {cP\over 2\pi} \sin^2\Theta,%
\label{r-theta}
\end{equation}
where $r=|{\bf r}|$, $\Theta$ is the polar angle, $P$ the rotation
period, $c$ the speed of light, and $\lambda$ a parameter which
characterizes the sorts of field lines ($\lambda \ga 1$,
$\lambda=1$ for the last open filed lines).
Based on Eq.(\ref{r-theta}), the angle $\Theta_\mu$ between the
direction of magnetic field at vector position {\bf r} and the
magnetic axis ${\bf \mu}$ reads,
\begin{equation}
\begin{array}{lll}
\cos\Theta_\mu&=&{2\lambda cP-6\pi r\over \sqrt{\lambda
cP(4\lambda cP-6\pi r)}},\\%
&=&{2-3\sin^2\Theta \over \sqrt{4-3\sin^2\Theta}}.%
\end{array}
\label{theta_mu}
\end{equation}
Therefore, assuming that elementary emission is along magnetic
field lines and according to Eq.(\ref{r-theta}-\ref{theta_mu})
and Appendix, we obtain the phase shift $\delta\phi$ as a function
of $r$
\begin{equation}
\delta\phi(r)={4\pi r\over \lambda cP} \sqrt{\lambda cP-2\pi r
\over 4\lambda cP - 6\pi r},%
\label{deltaphi}
\end{equation}
and the phase difference $\Delta\phi$ between two emission beam
components with heights $r_{\rm a}$ and $r_{\rm b}$ is
\begin{equation}
\Delta\phi = \delta\phi(r_{\rm a}) - \delta\phi(r_{\rm b}).
\label{Deltaphi}
\end{equation}

We note, from Eq.(\ref{deltaphi}), that $\delta\phi$ is only a
function of rotation period $P$ and distance $r$, but is not
relevant to magnetic inclination angle $\alpha$ and/or impact
angle $\beta$. The calculations of phase shift $\delta\phi$ for
pulsars with periods $P=0.03$s, 0.05s, 0.1s, 0.5s, and 1s are
shown in Fig.\ref{f.dphi-r}.
\vspace{0.2cm} \centerline{}
\centerline{\psfig{file=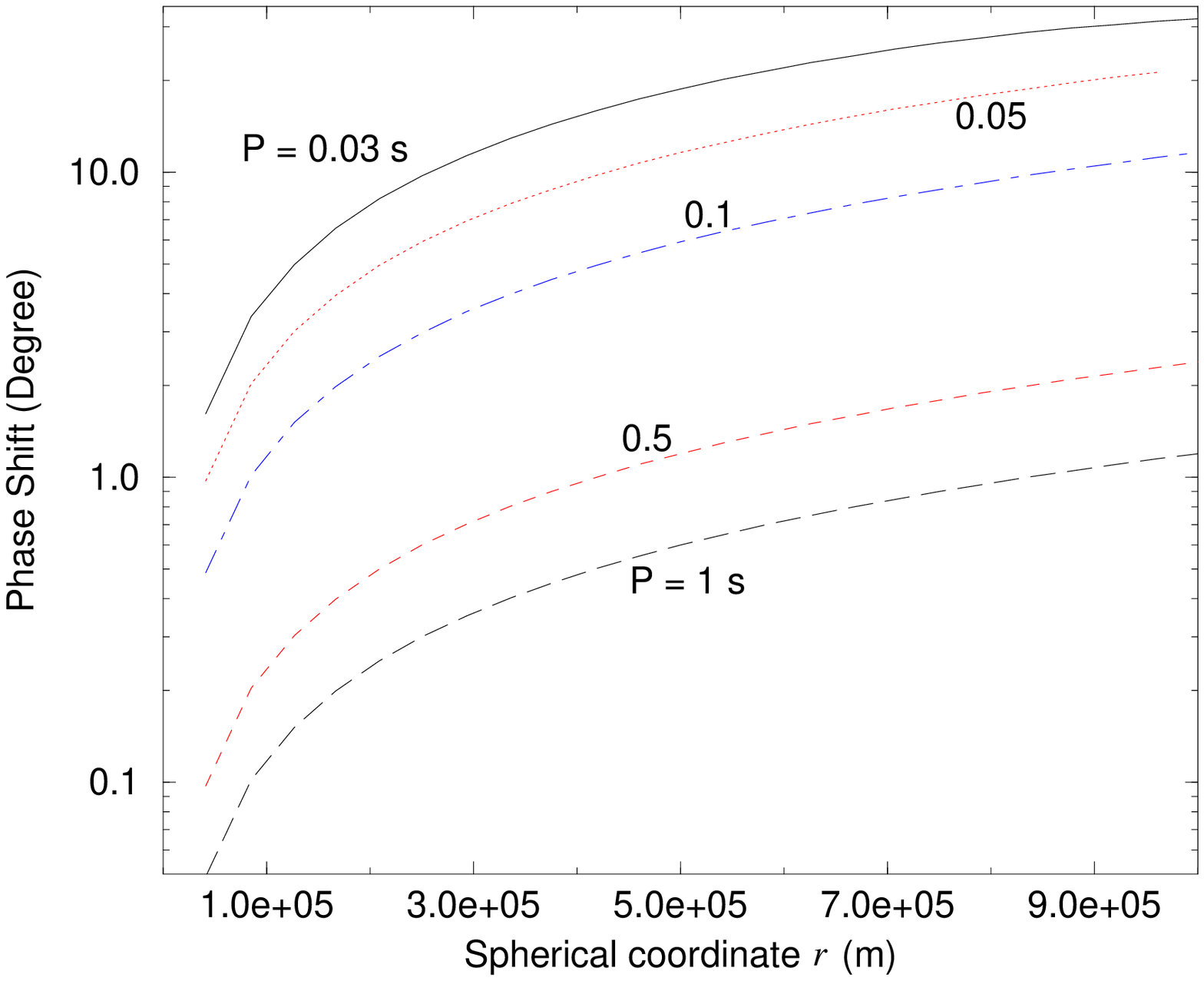,width=8cm,height=7cm}}
\figcaption{ The phase shifts $\delta\phi$ for pulsars with
$P=0.03$s, 0.05s, 0.1s, 0.5s, 1s, based on Eq.(\ref{deltaphi}).
The emission height (spherical coordinate $r$) is from 30km
to $10^3$km. $\lambda=1$.%
\label{f.dphi-r}}%
\vspace{0.2cm}

We see from Fig.\ref{f.dphi-r} that the retardation effect
arising from different emission heights is less important for
pulsars with longer periods $P$, but can not be negligible when
$P$ is small. For instance, the phase difference between
components with heights of 30km and 300km could be over $10^{\rm
o}$ if $P=30$ms. But the phase difference can only be about
$1^{\rm o}$ for emission components with heights 50km and 500km
if $P=0.5$s.

\subsection{The relationship between observed frequency and
shifted phase}

Theoretically, we can know the emission heights of beam components
at a certain observation frequency in the inverse Compton
scattering model$^{[\cite{ql98},\cite{xlhq00}]}$, based on which
the three components (core, inner cone, and outer cone) can be
understood naturally.
The frequency $\nu$ of radio wave emitted at position $r$ (or
height $r-R$, $R$ is the pulsar radius) can be gotten by following
Eq.(\ref{nu}-\ref{gamma}). Low frequency electromagnetic waves
with frequency $\nu_0$ are supposed to be produced near pulsar
surface due to RS-type vacuum gap sparking$^{[\cite{rs75}]}$.
These waves are assumed to propagate nearly freely in highly
inhomogeneous plasma in pulsar magnetospheres and inverse Compton
scattered by the secondary particles with Lorentz factor
$\gamma$, turning into radio waves observed with frequency
$\nu^{[\cite{xlhq00}]}$,
\begin{equation}
\nu = 1.5\gamma^2\nu_0(1-\sqrt{1-\gamma^{-2}}\cos\theta_{\rm i}),%
\label{nu}
\end{equation}
where the incident angle $\theta_{\rm i}$ is the angle between
the wave vector of low frequency wave and the direction of
electron moving (along magnetic fields in the approximation of
this paper), which can be calculated to be
\begin{equation}
\cos\theta_{\rm i}={2\cos\Theta+(R/r)(1-3\cos^2\Theta)
\over \sqrt{(1+3\cos^2\Theta)[1-2(R/r)\cos\Theta+(R/r)^2]}}.%
\label{thetai}
\end{equation}
The Lorentz factor of relativistic electron should decrease with
height\footnote{
Electrons will lost their kinetic energy, via, e.g., the curvature
radiation and/or the inverse Compton scattering of these electrons
off thermal X-ray photons from pulsar surface, when they are
moving along magnetic field lines.
},
which is suggested in the following form$^{[\cite{ql98}]}$
\begin{equation}
\gamma = \gamma_0\exp[-\xi{r-R\over R}],%
\label{gamma}
\end{equation}
where $\gamma_0$ is the initial Lorentz factor near pulsar
surface, $\xi$ a parameter reflecting the extend of the energy
loss.

\vspace{0.0cm}%
\centerline{}
\centerline{\psfig{file=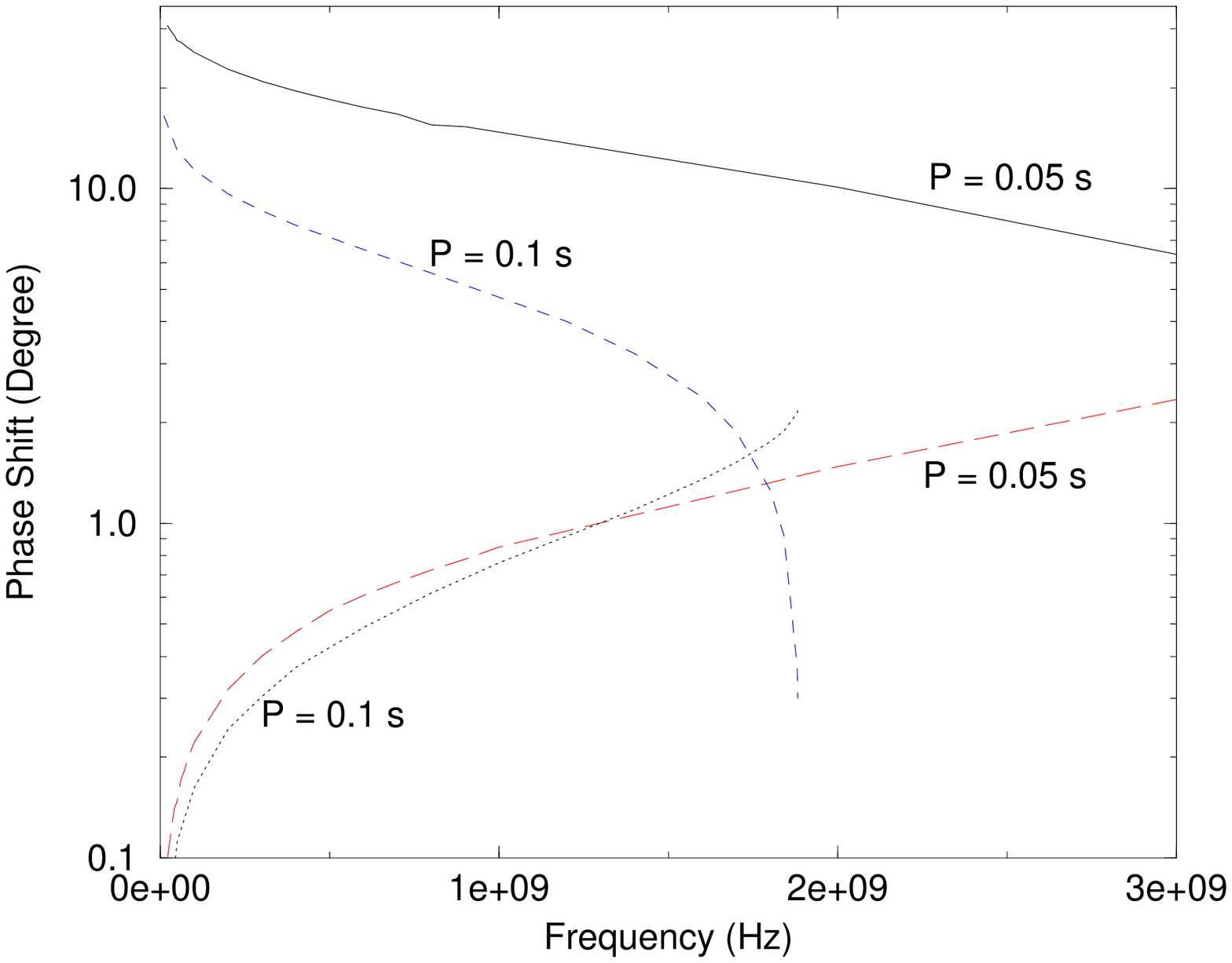,width=8cm,height=7cm}}
\figcaption{The phase differences between beam components, as
functions of observation frequency, in the ICS model. The upper
two lines (solid and short dashed) are the phase differences of
outer cone and inner cone, while others are of inner cone and
core. The pulsar periods are chosen to be 0.1s and 0.05s, and
other parameters to be $\gamma_0=3\times 10^3$, $\xi=3\times
10^{-2}$, $R=10^4$m, and $\nu_0=10^5{\rm H_Z}$. $\lambda=1$.
\label{f.phi_f}}%
\vspace{0.2cm}

The relation between observation frequency $\nu$ and position $r$
can be found from Eq.(\ref{r-theta}, \ref{nu}-\ref{gamma}). For a
certain $\nu$ we may find three solutions for $r$: $r_1,r_2,r_3$,
being the positions of core, inner cone, and outer cone,
respectively. The phase differences of beam components of outer
and inner cones, and of inner cone and core, are calculated for
observation frequency up-to a few GHz in the ICS model, which are
shown in Fig.\ref{f.phi_f}.

We see from Fig.\ref{f.phi_f} that, on one hand, at lower
frequency $\nu$, the phase differences $\Delta\phi_2\equiv
\delta\phi(r_3)-\delta\phi(r_2)$ of outer and inner cones are
notable. For example, $\Delta\phi_2=30^{\rm o}$ when $\nu=2\times
10^7$ Hz. $\Delta\phi_2$ decreases monotonously with the increase
of observation frequency $\nu$. On the other hand, the phase
differences $\Delta\phi_1\equiv \delta\phi(r_2)-\delta\phi(r_1)$
of core and inner cone components are small at lower $\nu$.
$\Delta\phi_1=0.05^{\rm o}$ contrary to $\Delta \phi_2$, $\Delta
\phi_1$ increases monotonously with the increase of $\nu$. The
observational tests and/or consequences of these features need
further investigations in the future.

\subsection{The possible distribution of singular points on the
celestial sphere}

There may be some points on the observational celestial sphere
where the linear polarization intensity $L=0$. We call such points
as ``singular points''. The polarization position angle should
jump\footnote{
The position angle may jump approximately 90$^{\rm o}$ if a line
of sight goes nearby a singular point.}
exactly 90$^{\rm o}$ when a line of sight goes across one of the
singular points$^{[\cite{xqh97,xq00}]}$. Therefore the
distribution of singular points on the celestial sphere would
determine the variation properties of position angle.

Generally, it is hard to calculate the singular point
distribution since there are many factors, being not well
understood, to cause depolarization$^{[\cite{xq00,l99}]}$ of
pulsar beamed radiation. However, if depolarization is mainly due
to the incoherent superposition of emission components at
different heights, and if there are only two components, we can
get the singular point distribution easily in this special case.

In this case the sufficient {\em and} necessary conditions that
should be satisfied at a singular point are
\begin{equation}
\left\{
\begin{array}{lll}
\psi_1-\psi_2 & = & \pm \pi/2\\
L_1 & = & L_2
\end{array},
\right.%
\label{s-n}
\end{equation}
where $\psi_{\rm i}$ and $L_{\rm i}$ (${\rm i}=1,2$ denotes the
first and second components, respectively) are the position
angles and linear polarization intensities of i$^{\rm th}$
components, respectively. The first condition, i.e.,
$\psi_1-\psi_2=\pm \pi/2$, is thus only a {\em necessary} one for
a singular point. Considering the identity
$\tan(\psi_1-\psi_2)={\tan\psi_1-\tan\psi_2\over
1+\tan\psi_1\tan\psi_2}$, we can re-write the necessary condition
below
\begin{equation}
\tan\psi_1\tan\psi_2=-1.%
\label{nec}
\end{equation}

The possible distribution of singular points on the celestial
sphere is calculated for this special case based on
Eq.(\ref{nec}), and is shown in Fig.\ref{f.sig}. We assume the
phase difference between the centers of the components due to
their different emission heights to be $\Delta\phi=15^{\rm o}$,
and the inclination angle $\alpha=$ 60$^{\rm o}$ or 30$^{\rm o}$
in our calculation. We find from Fig.\ref{f.sig} that the
distribution become ``flatter'' when the inclination angle
$\alpha$ is smaller, i.e., larger $\alpha$ may be more favorable
for a line of sight to go across a singular point, thus
increasing the possibility of position angle curve to ``jump''.

\vspace{0.0cm}%
\centerline{}
\centerline{\psfig{file=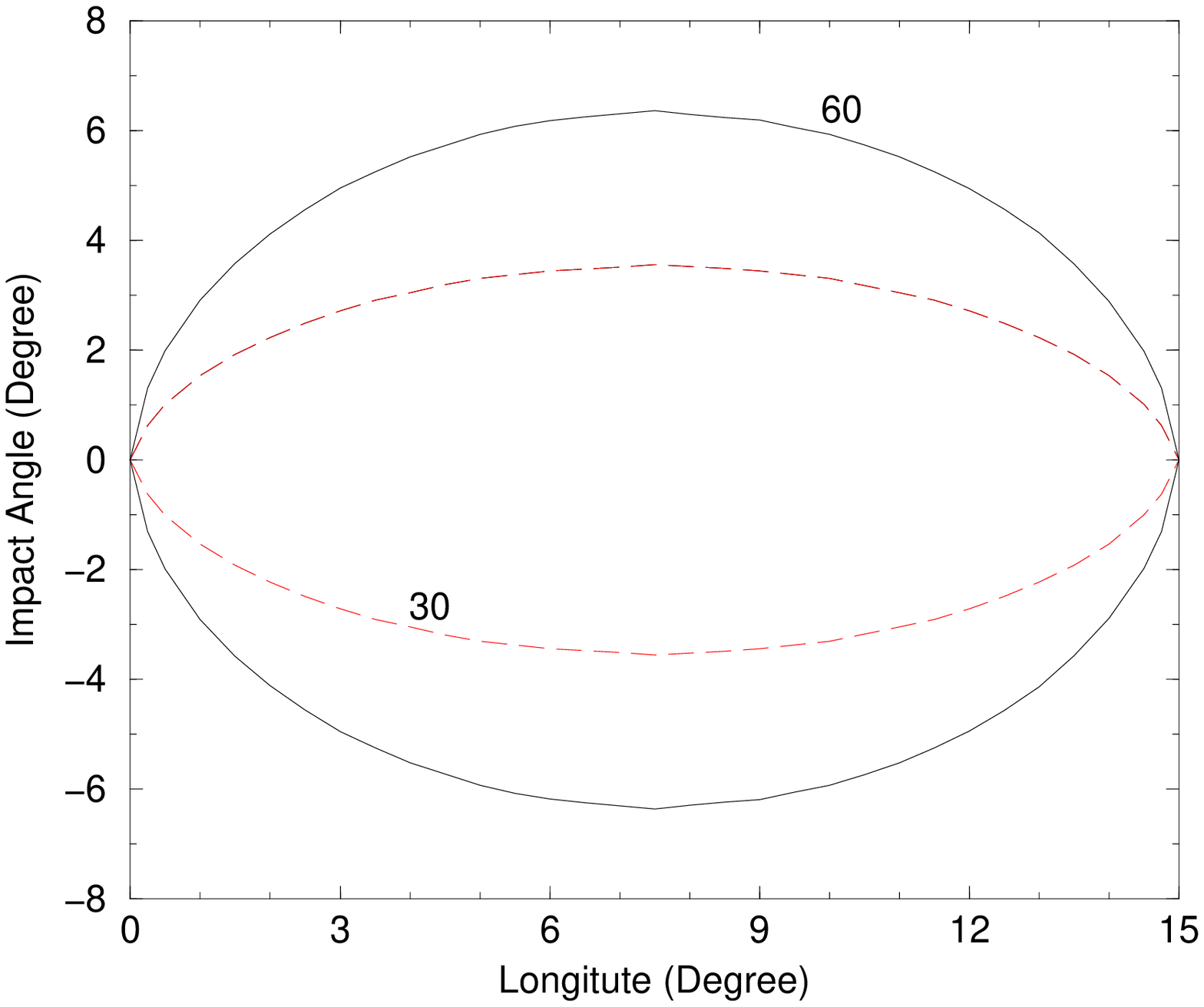,width=8cm,height=7cm}}%
\figcaption{A possible distribution of singular points on the
celestial sphere assuming that there are only two emission
components, and that the linear depolarization arises only from
the incoherent superposition of those two components. The phase
difference between the components is assumed to be 15$^{\rm o}$,
and the inclination angles are chosen to be 30$^{\rm o}$ and
60$^{\rm o}$, respectively, in the computation.%
\label{f.sig}}%
\vspace{0.2cm}

\section{Emission beams in the ICS model}

\vspace{0.0cm}%
\centerline{} \centerline{%
\psfig{file=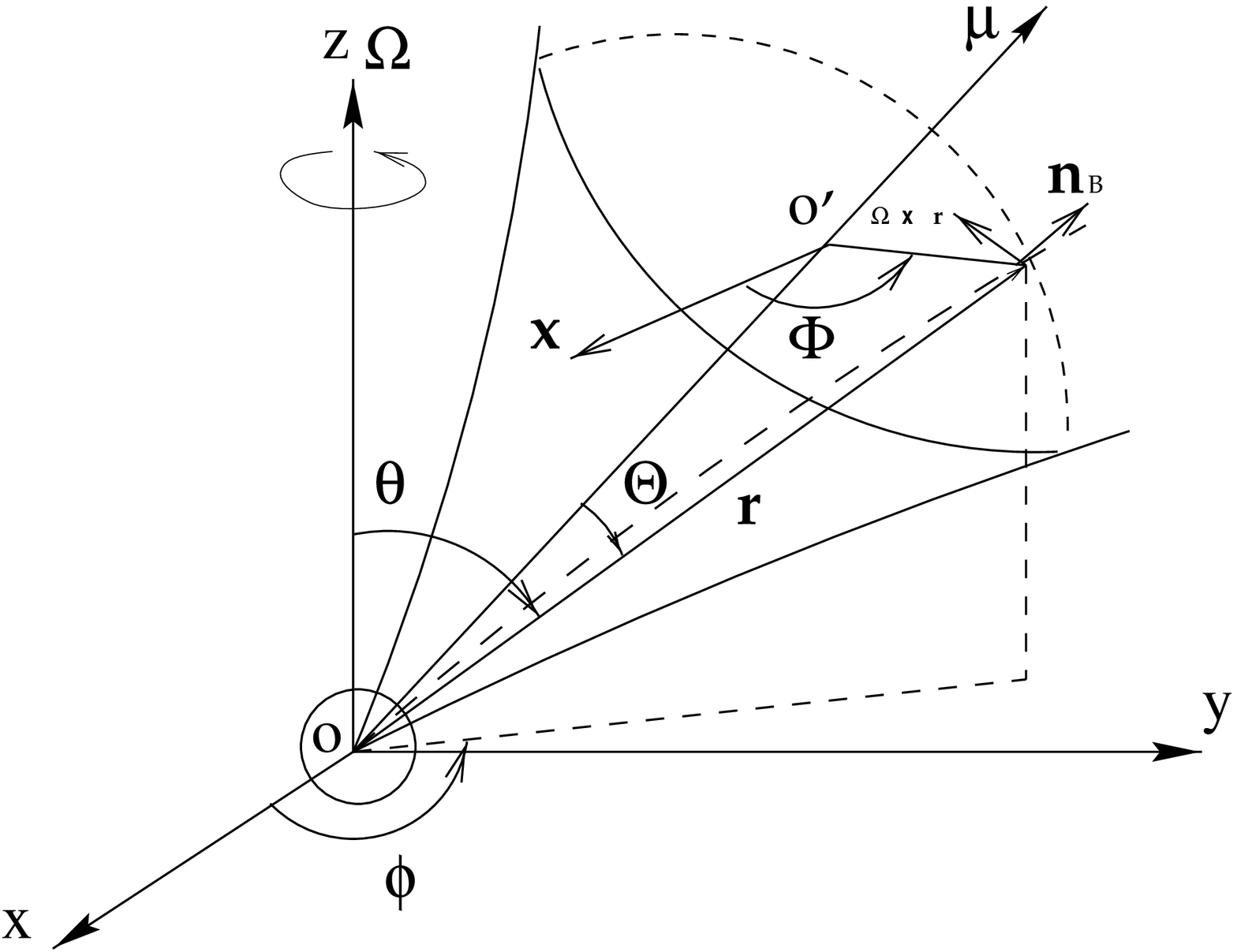,width=8cm,height=7cm}} %
\figcaption{A sketch of two coordinate systems. we assume the
$\mu$-coordinate, in which the magnetic axis $\mu$ is chosen as
the z-axis, turns clockwise an angle $\alpha$ (the inclination
angle) along the x-axis of $\Omega$-coordinate. A position vector
${\bf r}$ can be represented by $\{ r, \theta,\phi\ \}$ in
$\Omega$-coordinate, or by $\{ r, \Theta,\Phi\ \}$ in
$\mu$-coordinate. ${\bf n}_{\rm B}$ is the vector unit of magnetic
field.%
\label{f.rela}}%
\vspace{0.2cm}

In the above discussion, only the electron velocity along magnetic
field line (the poloidal velocity) is considered. However, as a
pulsar rotates, an electron does have a {\em toroidal} velocity
${\bf \Omega}\times {\bf r}$ (see Fig.\ref{f.rela}). Actually, the
velocity of an electron in the observer rest frame can be
obtained by the relativistic velocity transformation of the
poloidal velocity in the rotation frame. The inclusion of such
toroidal velocity has at least one implication: an emission unit
would shift forward a phase with respect to the magnetic direction
${\bf n}_{\rm B}$ since a relativistic electron emit photon at
the direction of its moving. In this section, we calculate the
beam shapes modified by inclusion the toroidal velocity, based on
the derived emission heights in the ICS model$^{[\cite{ql98}]}$.
In the following computation, two phase shift effects arising from
different emission heights and from toroidal velocity are
included. According to Fig.\ref{f.rela}, one can find the
relationship between $\{\theta, \phi \}$ and $\{\Theta, \Phi \}$,
\begin{equation}
\left\{
\begin{array}{lll}
\cos\theta & = & \cos\alpha\cos\Theta-\sin\alpha\sin\Theta\sin\Phi\\
\sin\phi & = & {\cos\alpha\sin\Phi\sin\Theta+\sin\alpha\cos\Theta
\over \sin\theta}\\
\cos\phi & = & {\cos\Phi\sin\Theta \over \sin\theta}
\end{array},
\right.%
\label{transf}
\end{equation}
which is useful below.

The linear rotation velocity at a position vector {\bf r} is $v$,
\begin{equation}
\begin{array}{lll}
v&=&|{\bf \Omega}\times {\bf r}|\\
&=& \Omega r\sqrt{1 - (\cos\alpha \cos\Theta -
\sin\alpha\sin\Theta\sin\Phi)^2}
\end{array}.
\label{v}
\end{equation}
Thus, by the relativistic velocity transformation, one finds two
components, $u_1$ (in the ${\bf \Omega}\times {\bf r}$ direction)
and $u_2$ (in the $[({\bf \Omega}\times {\bf r})\times {\bf
n}_{\bf B}]\times ({\bf \Omega}\times {\bf r})$ direction), of
the velocity of an electron in the observer rest frame,
\begin{equation}
\begin{array}{lll}
u_1&=&{-c\sqrt{1-\gamma^{-2}}\sin\alpha\sin\Theta\cos\Phi +
    v\sqrt{1+3\cos^2\Theta} \over %
    \sqrt{1+3\cos^2\Theta}-v\sqrt{1-\gamma^{-2}}
    \sin\alpha\sin\Theta\cos\Phi/c},\\
u_2&=&{c\sqrt{(1-\gamma^{-2})(1-v^2/c^2)(1+3\cos^2\Theta-
\sin^2\alpha\sin^2\Theta\cos^2\Phi)} \over %
    \sqrt{1+3\cos^2\Theta}-v\sqrt{1-\gamma^{-2}}
    \sin\alpha\sin\Theta\cos\Phi/c},
\end{array}
\label{vi}
\end{equation}
where $\gamma$ is the Lorentz factor of electron in the rotation
frame. Therefore the angle $\delta$ between ${\bf n}_{\rm B}$ and
the emitted photon is,
\begin{equation}
\delta\varphi = \cos^{-1}[{-\sin\alpha\sin\Theta\cos\Theta\over
\sqrt{1+3\cos^2\Theta}}]-\cos^{-1}[{u_1\over \sqrt{u_1^2+u_2^2}}], %
\label{delta}
\end{equation}
which is also the phase shift due to toroidal velocity.

Combining the phase shifts of $\delta\phi$ (Eq.(\ref{deltaphi}))
and $\delta\varphi$ (Eq.(\ref{delta})), we can calculate the
emission beams in the ICS model. We use the results in
Fig.\ref{f.phi_f} to decide the emission heights. The radio
emission is assumed to originated from a region between magnetic
field lines denoted by $\lambda=1$ and that by $\lambda=1.8$. The
radius of the stellar surface of the feet of the field lines
corresponding to $\lambda=1.8$ is $\sim \lambda^{-1/2}\theta_{\rm
p}\sim 0.75\theta_{\rm p}$, where $\theta_{\rm
p}=\sin^{-1}\sqrt{2\pi R\over cP}$ is the radius of polar cap.
The cases of rotation period $P=0.1 {\rm s}$ and $0.05$s, and
observation frequency $\nu=10^9$Hz are presented in Fig.\ref{f.1}
and Fig.\ref{f.5}. The parameters in the simulation are listed in
Table 1.

\begin{deluxetable}{lccccccccr}
\tablewidth{7.0in} \tablenum{1}%
\tablecaption{The parameters$^*$ chosen and calculated in the ICS
model for the  computations of emission beam components}%
\tablehead{ \colhead{$P$(s)} & \colhead{$\nu$(Hz)} &
\colhead{$r_1$(m)} & \colhead{$r_2$(m)} & \colhead{$r_3$(m)} &
\colhead{$\theta_{\mu 1}$} & \colhead{$\theta_{\mu 2}$} &
\colhead{$\theta_{\mu 3}$} & \colhead{$\Delta\phi_1$} &
\colhead{$\Delta\phi_2$} }
\startdata%
 0.1 & $10^9$ & 18820 & 81900 & 483820 & 5$^{\rm o}$.4 &
 11$^{\rm o}$.3 & 28$^{\rm o}$.1 & 0$^{\rm o}$.76 & 4$^{\rm o}$.74 \nl%
0.05 & $10^9$ & 20695 & 56040 & 697000 & 8$^{\rm o}$.0 &
 13$^{\rm o}$.2 & 50$^{\rm o}$.5 & 0$^{\rm o}$.85 & 14$^{\rm o}$.71 \nl%
\enddata
{$^*P$ - rotation period; $\nu$ - observation frequency; $r_{\rm
i}$ - emission height of component i; i$=1, 2, 3$ denotes core,
inner cone, and outer cone, respectively; $\theta_{\mu {\rm i}}$
- angular radius of component i; $\Delta\phi_1$ and
$\Delta\phi_2$ are the phase differences of core and inner cone,
and of inner and outer cones, respectively.}%
\vspace{0.2cm}
\end{deluxetable}{}

From Fig.\ref{f.1} and Fig.\ref{f.1}, we see the emission beams
are clearly {\em not} symmetric although the emission regions of
these components may be symmetric respect to the magnetic axes.
The width of the first pulse in a conal component would be much
smaller than that of the second one. Besides the clear phase
shifts between the components, the emission beams may not have
circular cross sections when the rotation period is small (see
Fig.\ref{f.5}). The pulses of impact angle $\beta>0$ arrive
earlier than that of $\beta<0$. This is understandable. Since the
emission region of $\beta>0$ is farther from the rotation axis
than that of $\beta<0$, the phase shift effect thus is more
remarkable due to larger toroidal velocity.

\vspace{0.0cm}%
\centerline{} \centerline{%
\psfig{file=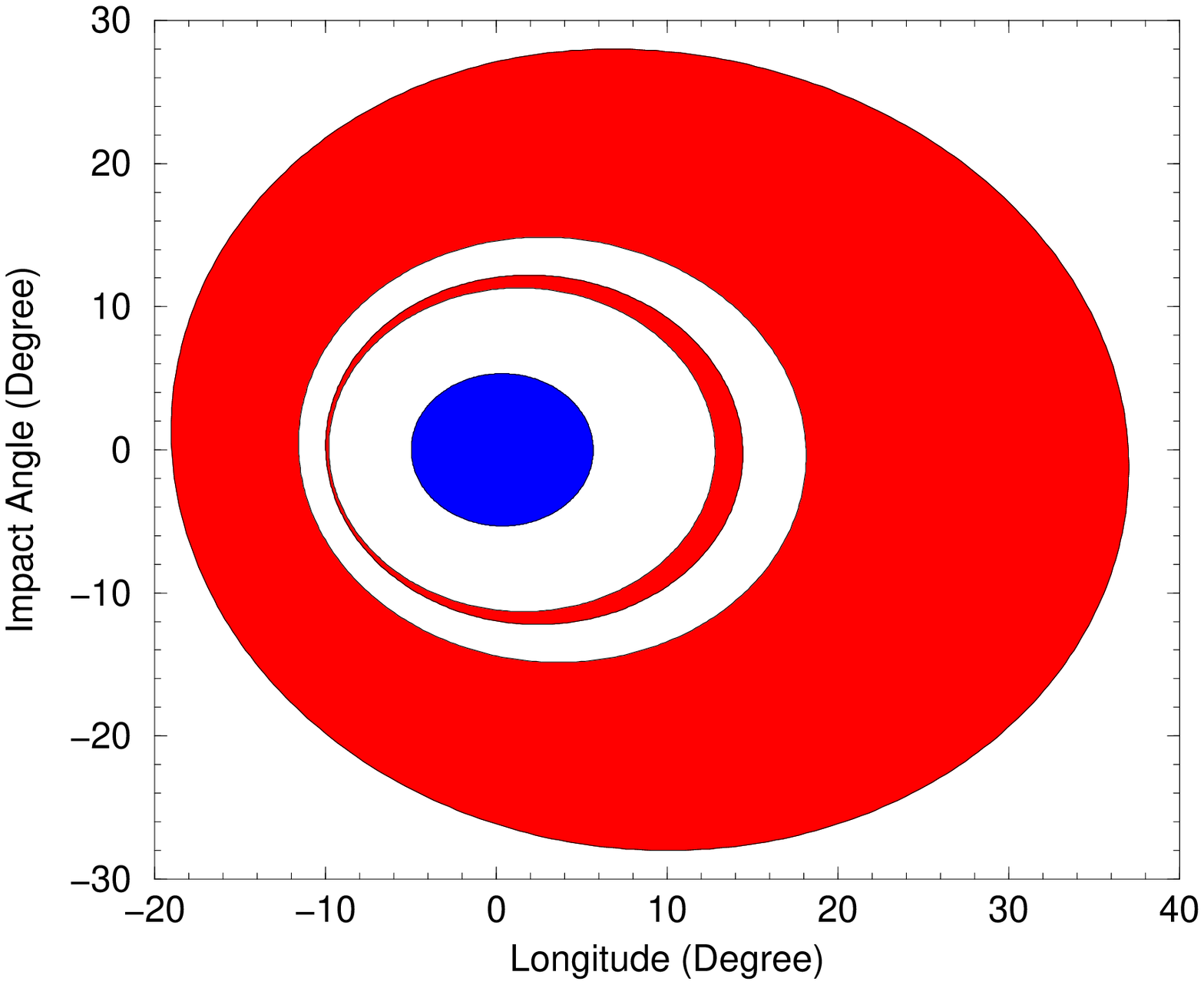,width=8cm,height=8cm}} %
\figcaption{The emission beams of core, inner cone, and outer cone
simulated in the ICS model, with the inclusion of the phase shift
effects both due to different emission heights and to toroidal
velocity. The inclination angle $\alpha=30^{\rm o}$, the rotation
period $P=0.1$s. Other parameters
are the same as in Fig.\ref{f.phi_f}. %
\label{f.1}}%

\vspace{0.0cm}%
\centerline{}%
\centerline{\psfig{file=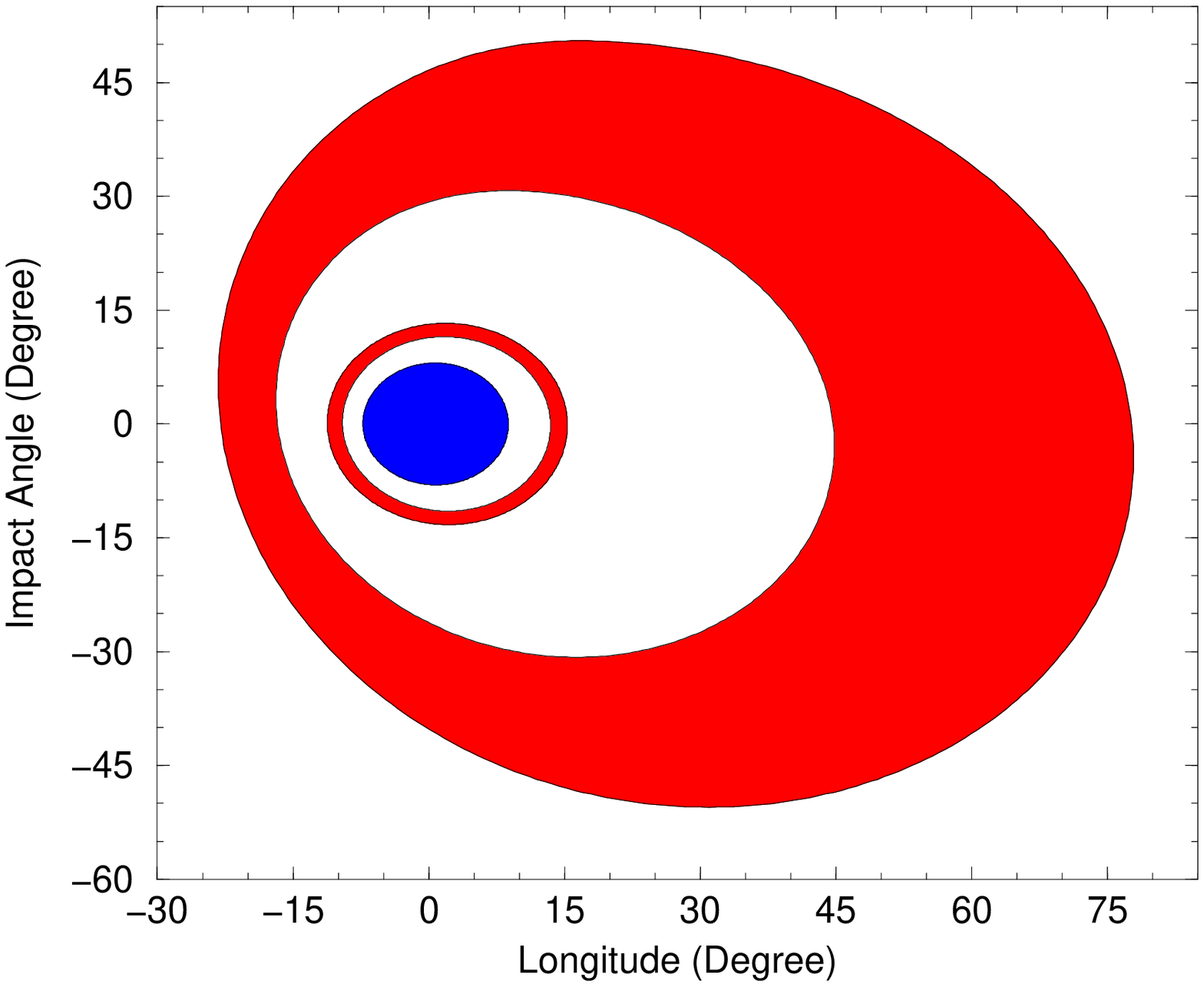,width=8cm,height=8cm}}%
\figcaption{Same as in Fig.\ref{f.1}, but for $P=0.05$s.%
\label{f.5}}%

\section{Conclusion and discussion}

A detail consideration of the relative longitude phase shift due
to different emission heights of three components is presented in
this paper. Our main conclusion can be summarized below.

1, An emission unit along direction ${\bf n}_1(\theta_0,\phi_1)$
at an arbitrary position ${\bf r}_1$ arrives earlier than another
emission unit along direction ${\bf n}_2(\theta_0,\phi_2)$ at
another position ${\bf r}_2$, the phase difference between those
two elements is $\phi_1-\phi_2+\Omega({\bf n}_1\cdot {\bf
r}_1-{\bf n}_2\cdot {\bf r}_2)/c$ ({\em if} this value is
positive), where $\Omega$ is the rotational angular velocity.

2, The phase shift effect is less important when the rotation
period $P>0.5$s.

3, The phase shift values, as a function of observation frequency
$\nu$ , are presented in the ICS model. As $\nu$ increases, the
shift between outer and inner cones decreases, while the shift
between inner cone and core
increases.

4, Larger inclination angles may be favorable for the appearance
of polarization position angle jumps.

6, The emission beams are not symmetric in the ICS model, which is
achieved by the inclusion of the phase shift effects both due to
different emission heights and to toroidal velocity (${\bf
\Omega}\times {\bf r}$).

In the simulation of the beam shapes of this paper, we just take
the symmetric emission region in the ICS model for simplicity.
However, the emission region should actually {\em not} be
symmetric respect to the magnetic axes when we consider that, for
a certain inverse Compton scattering process, the low-frequency
wave is emitted at an earlier sparking point, and that electrons
are not moving exactly along magnetic field lines. How about the
emission beams and polarization properties when these issues are
included? A further study is needed in the future.

\vspace{1cm}

\noindent %
{\bf acknowledgements:} We thank our pulsar group for discussions.

\begin{center}
\Large{{\bf Appendix}}
\end{center}

Observation shows that emission components (core, inner cone, and
outer cone) could come from different heights$^{[\cite{r83}]}$.
The longitude phase shifts between those components are thus
inevitable$^{[\cite{xqh97}]}$. We now try to determine this shift
value between two emission components. Our measures to deal with
this question is that:
first we find a virtual equivalent\footnote{%
We call two emission units are equivalent only if an observer at
infinity can not distinguish them, i.e., the observer detects
same intensity at same time for those two emission units.
}%
at origin for an emission unit at arbitrary position;
then we compare the phases of the virtual equivalents.
The phase difference between two virtual equivalents is actually
the phase difference observed of those two corresponding emission
units. Our conclusion can be expressed as following proposition:

\noindent
{\em An emission unit along a direction ${\bf
n}_0(\theta_0,\phi_0)$ at an arbitrary position ${\bf
r}(\theta,\phi)$ is equivalent observationally to a virtual
emission unit along direction ${\bf n}'_0(\theta_0,\phi'_0)$ at
origin, where $\phi'_0=\phi_0+\Omega \tau$, $\tau={{\bf r
}\cdot{\bf n}_0/c}$, $\Omega$ and $c$ are the rotational angular
velocity and the speed of light, respectively.}

\vspace{0.0cm} \centerline{}
\centerline{\psfig{file=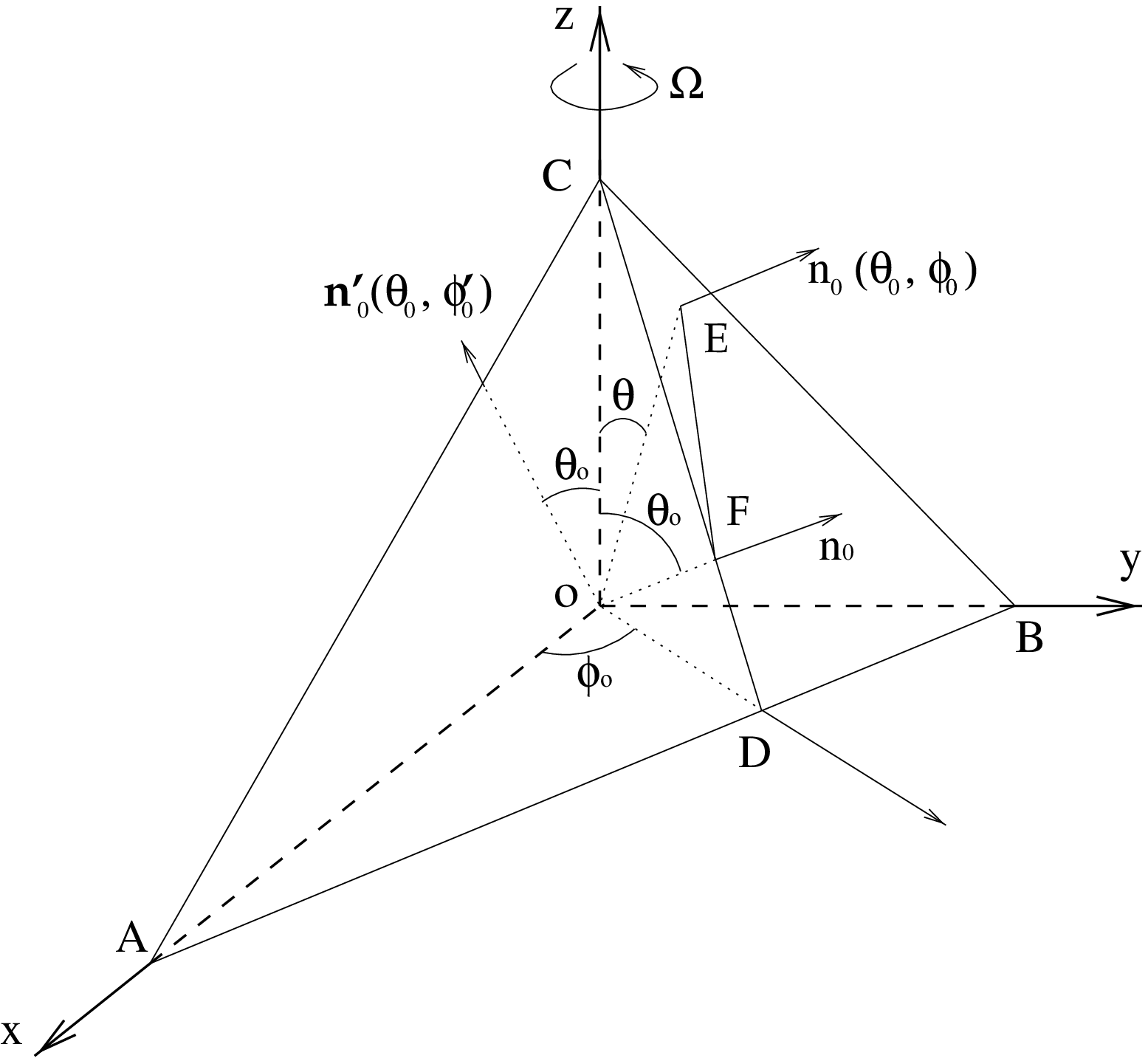,width=8cm,height=7cm}}
\figcaption{A sketch geometrical figure for proving the
conclusion to find an equivalent emission unit at origin. An
emission along ${\bf n}_0$ at point E (position vector ${\bf r}$)
is equivalent to the virtual one along ${\bf n}'_0$ at origin.
\label{Fig.A1}} %
\vspace{0.2cm}
We are to prove this below. As in Fig.\ref{Fig.A1}, we erect
coordinates o-xyz in the observation rest frame, where z-axis is
along rotational axis (${\bf \Omega}$), the origin is at the
center of pulsar. We are considering the emission along ${\bf
n}_0$ at an arbitrary point E. A plane ABC is created so that
${\bf n}_0$ is perpendicular to this plane. Because all emission
units along the direction ${\bf n}_0$ at those points which are
on the ABC plane should be observed simultaneously, the emission
along ${\bf n}_0$ at point\footnote{
Point F is the intersection of ABC plane and its perpendicular
line through point O.
} F is equivalent to the emission at point E if the intensities
are the same. We assume the time $t=0$ when a photon is emitted at
point E along ${\bf n}_0$. A photon emitted at point O along
${\bf n}_0$ when $t=-\tau$ ($\tau=|{\bf OF}|/c$) arrives at point
F when $t=0$ (still along ${\bf n}_0$). However, at $t=0$, the
emission at point O is along ${\bf n}'_0(\theta_0,\phi_0+\Omega
\tau)$ because of rotation. Therefore, an emission along ${\bf
n}_0$ at point E with position vector ${\bf r}$ is equivalent to
the virtual one along ${\bf n}'_0$ at origin O.

Now we calculate the value $\tau$. In the right-angled triangle
OEF, $|{\bf OF}| = {\bf r} \cdot {\bf n}_0$. We thus have $\tau =
{\bf r} \cdot {\bf n}_0/c$.

\vspace{1cm}

\noindent %

\end{document}